\documentclass[12pt]{article}
\usepackage[a4paper,text={16.8cm,22.4cm}]{geometry}
\usepackage{authblk}
\usepackage{amsmath,amsfonts,slashed,amssymb,tikz,bm,psfrag,graphicx,color,dsfont}
\usepackage{float}
\usepackage[colorlinks=true,linkcolor=red,citecolor=blue]{hyperref}
\usepackage{cite}
\allowdisplaybreaks
\begin{document}
\begin{titlepage}
\title{ Power Corrections to Pion Transition Form Factor in Perturbative QCD Approach }
\author[$a$]{Yue-Long Shen\footnote{shenylmeteor@ouc.edu.cn}}
\author[$b$]{Zhi-Tian Zou\footnote{zouzt@ytu.edu.cn}}
\author[$b$]{Ying Li\footnote{liying@ytu.edu.cn}}
\affil[$a$]{\it \small College of Information Science and
Engineering, Ocean University of China, Qingdao, Shandong 266100,
P.R. China} \affil[$b$]{\it \small Department of Physics, Yantai
University, Yantai 264005, China} \maketitle \vspace{0.2cm}
\begin{abstract}
In this paper we calculate the power corrections to the pion transition form factor within the
 framework of perturbative QCD approach on the basis of $k_T$ factorization.
 The power suppressed contributions from higher twist pion wave functions and
 the hadronic structure of photon are investigated. We find that there exists
 strong cancellation between the two kinds contributions, thus the total power
  corrections considered currently are very small, and the prediction of the leading power contribution
   with joint resummation improved perturbative QCD approach is almost unchanged.
    This result confirms that the pion transition form factor is a good platform to
    constrain the nonperturbative parameters in pion wave functions. Moreover,
    our result can accommodate the anomalous data from BaBar, or agrees with results from Belle according
    to the choice of Gegebauer moment in the pion wave function,
     and the more precise experimental data from Belle-II is expected.
\end{abstract}
\end{titlepage}
\section{Introduction}
\label{sect:Intro}
The pion-photon transition process  $\gamma^{\ast} \gamma \to \pi^{0}$ provides a golden place to test the strong interaction dynamics of hadronic reactions in the framework of QCD \cite{Lepage:1980fj,Efremov:1979qk,Duncan:1979ny,delAguila:1981nk,Braaten:1982yp,Kadantseva:1985kb,Melic:2002ij}.
The asymptotic and soft behaviors of the pion transition form factor $F_{\gamma^{\ast} \gamma \to \pi^0} (Q^2)$ have been already given as
\begin{eqnarray}
\lim_{Q^2\to\infty}Q^2 F_{\gamma^{\ast} \gamma \to \pi^0} (Q^2)&=&\sqrt{2} f_\pi=0.185,\label{eq-1}\\
\lim_{Q^2\to 0}F_{\gamma^{\ast} \gamma \to \pi^0} (Q^2)&=&\frac{\sqrt{2}}{4\pi^2f_\pi},\label{eq-2}
\end{eqnarray}
where $Q^2$ stands for the momentum transfer squared carried by the virtual photon, and the pion decay constant is $f_\pi= 0.131\mathrm{GeV}$. The former had been predicted within  perturbative QCD (PQCD) in the collinear factorization theorem, and the latter one could be determined from the axial anomaly in the chiral limit. In 2009, the experimental result of BaBar \cite{Aubert:2009mc} on $F_{\gamma^{\ast} \gamma \to \pi^0} (Q^2)$ exhibits an intriguing dependence on $Q^2$: for $Q^2>10 \mathrm{GeV}^2$, $F_{\gamma^{\ast} \gamma \to \pi^0} (Q^2)$ lies above Eq.(\ref{eq-1}) and continues to grow up to  $Q^2\approx 40 \mathrm{GeV}^2$. In contrast to BaBar, Belle also presented their measurements in the region $4 \mathrm{GeV}^2 \leq Q^2 \leq 40 \mathrm{GeV}^2$ \cite{Uehara:2012ag}, and  the rapid growth in the higher $Q^2$ region does not appear. Since there is no final confirmation on this discrepancy, various phenomenological approaches as well as lattice QCD simulations (see for instance \cite{Masjuan:2012wy, Hoferichter:2014vra, Gerardin:2016cqj}) have been employed to explain the data.

To accommodate the anomalous BaBar data at high $Q^2$, one approach is to introduce an ``exotic" twist-two pion light-cone distribution amplitude (LCDA) with the non-vanishing end-point behavior \cite{Radyushkin:2009zg,Polyakov:2009je}, but it was found to be equivalent to the introduction of a sizable nonperturbative soft correction from the transverse momentum dependent(TMD) pion wave function \cite{Agaev:2010aq}. Actually, in the framework of $k_T$ factorization \cite{Li:1992nu}, the ``exotic" wave function is not necessary \cite{LM09}. At leading power the pion transition form factor has  been studied with perturbative QCD(PQCD) approach based on $k_T$ factorization  at one-loop level \cite{Nandi:2007qx, Musatov:1997pu, Wu:2010zc}, and the resummation of the large TMD logarithms and threshold logarithms lead to the Sudakov factor and jet function, and an appropriate parameterization of the latter one can be used to explain the Babar data. To avoid light-cone singularity in the TMD wave function, an off-light-cone vector should be included in the definition of the wave function \cite{Co03,CS08}, then additional rapidity logarithms arise\footnote{Actually there is additional self-energy divergence attributed to the infinitely long dipolar Wilson lines existing \cite{BBDM} after the off-light-cone vector is adopted, and a more complicated definition of TMD wave function with the dipolar Wilson links and the complicated soft subtraction \cite{Collins:2011zzd}, or with non-dipolar off-light-cone Wilson links \cite{Li:2014xda}, is required. The definition of the wave function is essential in the soft subtraction, while for phenomenological application, it is more important to extract the hard kernel, and the traditional definition is still adopted in the present paper.}. Taking advantage of joint resummation technique one can resum  the large logarithms $\ln^2{{k_{\perp}^2} /Q^2}$ and $\ln^2 x$ and $\ln \zeta^2$ simultaneously \cite{Li:2013xna}. Using the joint resummation improved factorization formula, the BaBar data can also be explained if an appropriate Gegenbauer meoment of pion is employed.

In the PQCD framework based on $k_T$ factorization, though the next-leading order of $\alpha_s$ correction at the leading power contribution to this process has been studied \cite{LM09,Nandi:2007qx}, the higher power corrections have not been investigated till now. In fact, the scaling violation implied by the BaBar data \cite{Aubert:2009mc} also indicates the importance of subleading power corrections. The next-to-leading power(NLP) effects have been extensively studied in the collinear factorization framework, the contribution from higher twist pion LCDAs \cite{Khodjamirian:1997tk}, the hadronic structure of photon \cite{Wang:2017ijn} were considered, and in \cite{Agaev:2010aq,Agaev:2012tm}, the soft correction to the leading twist effect is evaluated with the dispersion approach and found to be crucial to suppress the contributions from higher Gegenbauer moments of the twist-2 pion LCDA \cite{Kroll:2010bf,Li:2013xna}. Power suppressed contributions in collinear factorization in general suffer from endpoint singularity and factorization breaks down. Alternative approaches such as dispersion approach need to be employed, but large uncertainty arises when quark hadron duality assumption is employed. It is well known to us that in PQCD approach the transverse momentum of parton can regularize the endpoint singularity so that the factorization is expected to work at NLP. Since the NLP corrections have not been studied previously, the aim of this article is to study the its contributions to the pion transition form factors within the PQCD approach that is based on $k_T$ factorization. In the current work, we shall consider two kinds of power suppressed contributions, which are from the higher twist pion wave functions and hadronic structure of photon, respectively.

The outline of this paper is  as follows:  in section \ref{sect:lp} we review the $k_T$ factorization and the joint resummation of $F_{\gamma^{\ast} \gamma \to \pi^0} (Q^2)$ at the leading power. In Sec.\ref{sect:higher-power}, the analytic calculation of the power suppressed contributions will be presented. The numerical results and discussions are given in section \ref{numerical}. We will summarize this work in the last section.
\section{Factorization and Resummation at Leading Power}
\label{sect:lp}
The pion transition form factor is defined via the matrix element
\begin{eqnarray}
\langle \pi(p) | j_{\mu}^{\rm em}  | \gamma (p^{\prime}) \rangle =
g_{\rm em}^2 \, \epsilon_{\mu \nu \alpha \beta} \, q^{\alpha} \,
p^{\beta} \, \epsilon^{\nu}(p^{\prime}) F_{\gamma^{\ast} \gamma
\to \pi^0} (Q^2) \,,
\end{eqnarray}
where $q=p-p^{\prime}$, $p$ and $p'$ refer to the four-momentum of the pion and the on-shell photon respectively,  the electro-magnetic current
\begin{eqnarray}
j_{\mu}^{\rm em} = \sum_q \, g_{\rm em} \, Q_q \, \bar q \,
\gamma_{\mu} \, q \, \,. \label{em current definition}
\end{eqnarray}

In $k_T$ factorization framework, the pion transition form factor is factorized into the convolution of the TMD wave function of pion and the hard kernel. The $k_T$ factorization theorem can be derived diagrammatically \cite{Nagashima:2002ia} by applying the eikonal approximation to collinear particles and the Ward identity to the diagram summation in the leading infrared regions. The TMD wave function of pion is defined by
\begin{eqnarray}
\Phi(u,k_{T}, \zeta^2, \mu_f)&=&\int\frac{dy^+}{2\pi }\frac{d^2y_T}{(2\pi)^2}e^{-i up^- y^{+} +i {\bf k}_{T}\cdot {\bf y}_T} \nonumber \\
&& \times\langle 0 |{\bar q}(y) W_y(v)^{\dag} \, I_{v;y,0} \,W_0(v) \not\! n_+\gamma_5 q(0)| \pi (p)  \rangle\,, \label{de1}
\end{eqnarray}
where $\mu_f$ is the factorization scale, the coordinate  $y=(y^+,0,{\bf y}_{T})$, and $u p^-$ and ${\bf k}_{T}$ are the longitudinal and transverse momenta carried by the anti-quark $\bar q$, respectively. The Wilson line
\begin{eqnarray}
\label{eq:WL.def} W_y(v) = {\cal P} \exp\left[-ig \int_0^\infty \,
d\lambda \, v \cdot A(y+\lambda v)\right],
\end{eqnarray}
has been introduced for maintaining the gauge invariance, where $\cal P$ denotes the path-ordered exponential. The avoid light-cone singularity, the non-light-like vector $v$ is employed \cite{Co03}. The transverse gauge link $I_{v;y,0}$ does not contribute in the covariant gauge \cite{CS08}.

For  pion transition form factor, the next-to-leading order(NLO) hard kernel has been calculated in \cite{Nandi:2007qx}. The  non-light-like vector $v$ will give rise to a new rapidity parameter $\zeta^2=\frac{4(n_{-} \cdot v)^2} {v^2}$ in both the NLO corrections to  wave function and hard kernel. For the rapidity parameter $\zeta^2=2$, the $k_T$ factorization formula at leading power of $1/Q^2$ under the conventional resummations was given by \cite{LM09}
\begin{eqnarray}
F(Q^2) &=&  \frac{\sqrt{2} \, f_{\pi}}{3}\int_0^1 du
\int_0^{\infty} b \, db \, \overline{\Phi}(u, b,t)
\,\, e^{-S(u,b,Q,t)} \,\, S_t(u,Q)   \,  \nonumber \\
&& \times K_0(\sqrt{u} Q \, b)   \left [1- \frac{\alpha_s(t) C_F}{4 \pi} \left ( 3 \ln { \frac{t^2b }{2 \sqrt{u} Q}} + \gamma_E + 2
\ln u+ 3 -\frac{\pi^2}{3}\right ) \right ]  \,,
\label{conventional PQCD}
\end{eqnarray}
where $t$ is the factorization scale. The Sudakov factor $S(u,b,Q,t)$ sums the double logarithm $\ln^2 ({k_T^2 / Q^2})$ and the single logarithm $\ln ({t^2 / Q^2})$, where the impact-parameter $\bf b$ is conjugated to the transverse momentum, and it is more convenient to resum large logarithms in $\bf b$-space than in transverse momentum space. The threshold factor from the resummation of $\ln^2 u$ has been parameterized as
\begin{eqnarray}
S_t(u,Q) &=&\frac{2^{1+c(Q^2)} \, \Gamma (\frac{3}{2}+c(Q^2))}{
\sqrt{\pi} \, \Gamma(1+c(Q^2)) } \, \left [ u (1-u) \right
]^{c(Q^2)}   \,,
\end{eqnarray}
It was found that the nontrivial $Q^2$ dependence of the factor $c(Q^2)$ is important in the explanation of BaBar data \cite{LM09}. Since the NLO QCD corrections will generate the mixed logarithm $\ln u\ln(\zeta^2P^{-2}/k_T^2)$ in both the pion wave function and the hard kernel, the double logarithms need to be resummed. In \cite{Li:2013xna},  an evolution equation has been constructed to resum the mixed logarithm $\ln u\,\ln(\zeta_P^2/k_T^2)$. It is more convenient to perform the resummation in the moment and impact parameter space, and the result reads
\begin{eqnarray}
\tilde{\Phi}(N, b, \zeta^2, \mu_f)&=&{\rm exp} \bigg \{ -
\int_{\zeta_0^2}^{\zeta^2} \frac{d \tilde{\zeta}^2 }{\tilde{\zeta}^2
} \, \left [\int_{\mu_0(\tilde{\zeta})}^{\mu_1(\tilde{\zeta})} \,
\frac{d \tilde{\mu}}{ \tilde{\mu}} \, \lambda_K(\tilde{\mu}) \,
\theta \left ( \mu_1(\tilde{\zeta}) - \mu_0(\tilde{\zeta})  \right
) \right ]
\nonumber \\
&&  \hspace{1.0 cm}  + {3 \over 2} \, \int_{\mu_i}^{\mu_f} \, {d
\tilde{\mu} \over \tilde{\mu}} \, {\alpha_s(\tilde{\mu}) \, C_F
\over  \pi}  \bigg \} \hspace{0.2 cm}  \tilde{\Phi}(N, b,
\zeta_0^2, \mu_i) \,, \\
\tilde{H}(N, b, \zeta^2, Q^2,
\mu_f)&=&{\rm exp} \bigg \{\int_{\zeta^2}^{\zeta_1^2} {d
\tilde{\zeta}^2 \over\tilde{\zeta}^2 } \, \left
[\int_{\mu_0(\tilde{\zeta})}^{\mu_1(\tilde{\zeta})} \, {d
\tilde{\mu} \over \tilde{\mu}} \, \lambda_K(\tilde{\mu}) \, \theta
\left ( \mu_1(\tilde{\zeta}) - \mu_0(\tilde{\zeta})  \right )
 \right ]
\nonumber \\
&&  \hspace{1.0 cm}  - {3 \over 2} \, \int^{\mu_f}_{t} \, {d
\tilde{\mu} \over \tilde{\mu}} \, {\alpha_s(\tilde{\mu}) \, C_F
\over  \pi}  \bigg \} \hspace{0.2 cm}  \tilde{H}(N, b, \zeta_1^2,
Q^2, t) \label{resummation of pion wavefunction}
\end{eqnarray}
where $\lambda_{{K}}={\alpha_s \, C_F \over 2 \pi} \,$, $\mu_i$ is the initial scale of the RG evolution. The bounds $\zeta_0^2$ and $\zeta_1^2$ are chosen in order to eliminate the large logarithms in the initial conditions of the pion wave function and the hard kernel. The joint-resummation improved pion wave function modifies both the longitudinal and transverse momentum distributions, and both the small $u$ and $b$ regions are more highlighted after resummation. In the pion transition form factor, by choosing appropriate $\zeta_1$, the hard kernel without large double logarithms reads
\begin{eqnarray}
H^{(1)}(u,k_T, \zeta_1^2,  Q^2, t)&=& -{\alpha_s(t) C_F \over 4
\pi} \bigg ( 3 \ln {t^2 \over  u Q^2+k_T^2} + \ln 2 +2  \bigg )
\,.
\end{eqnarray}
If a specific model of pion wave function has been employed, the resummation improved wave function can be transformed to the momentum space, and the joint-resummation improved factorization formula for the pion transition form factor is obtained as
\begin{eqnarray}
F(Q^2) &=& {\sqrt{2} \, f_{\pi} \over 3} \int_0^1 du
\int_0^{\infty} b \, db \, \overline{\Phi}(u, b, \zeta_1^2, t)
K_0(\sqrt{u} Q \, b)  \,  \nonumber \\
&& \times\left [1-  {\alpha_s(t) C_F \over 4 \pi} \left ( 3 \ln {
t^2 \, b \over 2 \sqrt{u} Q} + \ln 2 + 2   \right ) \right ]   \,,
\label{joint resummation PQCD}
\end{eqnarray}
where the expression the joint resummation improved wave function $\overline{\Phi}(u, b, \zeta_1^2, t)$ can be found in the ref.\cite{Li:2013xna}. In Eq. (\ref{joint resummation PQCD}) all the large logarithmic terms are collected by the resummed wave function, and the hard kernel is free from large logarithms.

\section{Subleading Power Corrections} \label{sect:higher-power}
In this section we aim at investigating the subleading power corrections to the pion transition form factor. As the transverse momentum of the parton of the pion meson is kept in the PQCD approach, the endpoint singularity does not appear, thus we can still take advantage of factorization method to evaluate the power suppressed contributions. As claimed in \cite{Agaev:2010aq}, a power-like falloff of the form factor  in the large-$Q^2$ limit can be generated by both ``direct" photon and  ``hadronic" photon contributions. The former one indicates that the hard subgraph includes both photon vertices, which starts from leading power ${\cal O}(1/Q^2)$, while the higher twist pion wave functions can give power suppressed contribution; the latter  is at most ${\cal O}(1/Q^4)$.  In the following we will consider two kinds of subleading power corrections within PQCD framework, that is from higher twist pion wave functions and the hadronic structure of photon.

\subsection{Higher-twist pion wave functions}
To evaluate the contribution from higher twist pion wave functions, firstly the definition of higher twist TMD wave functions similar to Eq. (\ref{de1}) are required. For simplicity, we assume that the initial pion wave function  can be factorized into the longitudinal and transverse parts,
\begin{eqnarray}
\Phi_i(u,k_T, \zeta_0^2, \mu_i)= \phi_i(u, \zeta_0^2, \mu_i) \, \,
\Sigma(k^2_T) \,, \label{initial condition: general}
\end{eqnarray}
For definiteness, the transverse momentum distribution is taken as
\begin{eqnarray}
\Sigma(k^2_T)= 4 \pi \beta^2 \exp(-\beta^2 \, k_T^2) \,,
\end{eqnarray}
where the prefactor is introduced to obey the normalization
\begin{eqnarray}
\int \frac{d^2 k_T}{(2\pi)^2 } \, \Sigma(k^2_T)=1\,.
\end{eqnarray}
The longitudinal momentum distribution $\phi_i(x, \zeta_0^2, \mu_i)$ is assumed to be the same as the LCDA. For the two-particle pion LCDAs, we have
\begin{eqnarray}
  \kappa_q\langle \pi(p)|\bar q(y) \gamma_\mu \gamma_5 q(0)|0\rangle
  &=& - i  f_\pi \,p _\mu \, \int_0^1 du \,  e^{i u \, p\cdot y }\, [\phi(u)+y^2g_1(u)]
\nonumber \\
&+&   f_\pi \,(y_\mu-{y^2p_\mu\over p\cdot y}) \, \int_0^1 du \,
e^{i u \, p\cdot y }\, g_2(u),
\nonumber \\
  \kappa_q\langle \pi(p)|\bar q(y) i \gamma_5 q(0)|0\rangle
  &=& f_\pi \mu_\pi \,  \int_0^1 du \,
  e^{i u \, p\cdot y } \, \phi_p(u),
\nonumber \\
  \kappa_q\langle \pi(p)|\bar q(y) \sigma_{\mu\nu}\gamma_5 q(0) |0\rangle
  &=& i f_\pi \mu_\pi \, (p_\mu y_\nu - p_\nu y_\mu) \,
      \int_0^1 du \, e^{i u \, p\cdot y } \,
      \frac{\phi_\sigma(u)}{6},
\label{pidadef}
\end{eqnarray}
where $\kappa_u=-\kappa_d=1/\sqrt{2}$ for $\pi_0$ meson. $\phi_\pi(u)$ is twist-2, $\phi^\pi_P(u)$ and $\phi^\pi_\sigma(u)$ are twist-3, and $g^\pi_1(u),g^\pi_2(u)$ are twist-4.

The three-particle pion LCDAs are also defined by
\cite{Khodjamirian:1997tk}
\begin{eqnarray}
&&\kappa_q\langle \pi(p)|\bar q(y) \gamma_\mu \gamma_5
g_sG_{\alpha\beta}(vy)q(0)|0\rangle =f_\pi \left(p_\beta
g_{\alpha\mu}-p_\alpha g_{\beta\mu}-{y_\alpha p_\beta -y_\beta
p_\alpha
  \over p\cdot y}p_\mu\right)
  \nonumber \\
  &&\times\int_0^1 [D\alpha_i] \, \phi^\pi_\perp(\alpha_i) e^{i  \, p\cdot y(\alpha_q+v\alpha_g) }\,
+  f_\pi {y_\alpha p_\beta -y_\beta p_\alpha
  \over p\cdot y}p_\mu
  \int_0^1 [D\alpha_i] \, \phi^\pi_\parallel(\alpha_i) e^{i  \, p\cdot y(\alpha_q+v\alpha_g) }\,,\nonumber \\
&&\kappa_q\langle \pi(p)|\bar q(y) \gamma_\mu \gamma_5 g_s\tilde
G_{\alpha\beta}(vy)q(0)|0\rangle=  if_\pi \left(p_\beta
g_{\alpha\mu}-p_\alpha g_{\beta\mu}-{y_\alpha p_\beta -y_\beta
p_\alpha
  \over p\cdot y}p_\mu\right)
\nonumber \\
&\times&   \int_0^1 [D\alpha_i] \, \tilde\phi^\pi_\perp(\alpha_i)
e^{i \, p\cdot y(\alpha_q+v\alpha_g) }\, + if_\pi {y_\alpha
p_\beta -y_\beta p_\alpha
  \over p\cdot y}p_\mu
  \int_0^1 [D\alpha_i] \, \tilde\phi^\pi_\parallel(\alpha_i) e^{i  \, p\cdot y(\alpha_q+v\alpha_g)
  },
\end{eqnarray}
here we have employed the following notations for the dual field strength tensor and the integration measure
\begin{eqnarray}
\widetilde{G}_{\alpha \beta}&=& {1 \over 2} \, \varepsilon_{\alpha
\beta \rho \tau }  \, G^{\rho \tau} \,, \qquad \int [{\cal D}
\alpha_i] \equiv \int_0^1 d \alpha_q \, \int_0^1 d \alpha_{\bar q}
\, \int_0^1 d \alpha_g \, \delta \left (1-\alpha_q - \alpha_{\bar
q} -\alpha_g \right ),
\end{eqnarray}
we note that all three-particle LCDAs are twist-4.

\begin{figure}
\begin{center}
\includegraphics[width=0.46 \columnwidth]{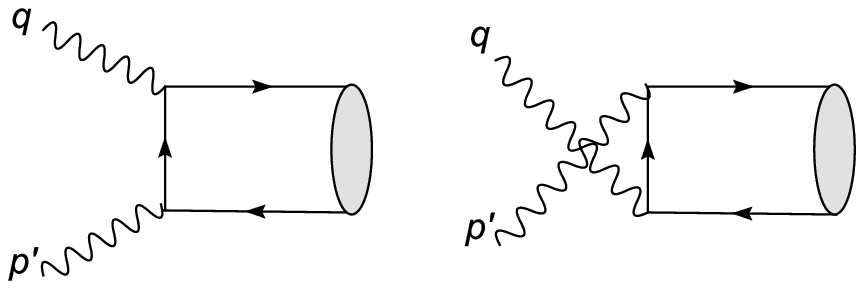} \hspace{1.0 cm}
\includegraphics[width=0.46 \columnwidth]{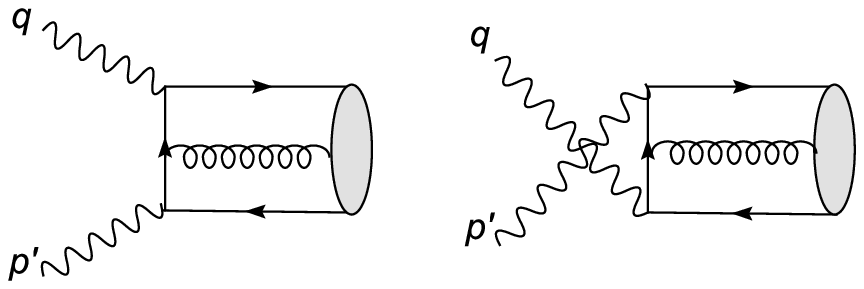} \\
(a)  \hspace{7 cm}    (b) \\
\vspace*{0.1cm} \caption{Fenynman diagrams of contribution of the
pion transition form factor from two-particle and 3-particle wave
functions} \label{fig:feyn1}
\end{center}
\end{figure}

It is straightforward to obtain the factorization formula for two-particle twist-4 contribution through evaluating the Feynman diagrams Fig. \ref{fig:feyn1}a,
\begin{eqnarray}
 F^{2PT4}(Q^2)
 &=&-{2\sqrt{2}f_\pi \over 3Q}\int_0^1 {du\over\sqrt{u}}\int_0^\infty
 b^2dbK_1(\sqrt{u}Qb)[g_1(u,b)+G_2(u,b)],
 \end{eqnarray}
where $G_2(u)=-\int_0^ug_2(v)$, and both the wave functions and hard kernel have been transformed into the impact parameter space.

To evaluate three-particle LCDA contribution, we need to keep the one-gluon part for the light-cone expansion of the quark propagator in the background gluon field
\begin{eqnarray}
 \langle 0|T\{{q}(y),\bar{q}(0)\}|0\rangle_{G}&\supset& i\int\frac{d^{4}k}{(2\pi)^{4}}
 e^{-ik\cdot y}\int_{0}^{1}dv[\frac{vy_{\mu}\gamma_{\nu}}{k^{2}}
 -\frac{\not \! k\sigma_{\mu\nu}}{2k^4}]G^{\mu\nu}(vy)\nonumber
\end{eqnarray}where
 $ G^{\mu\nu}=i[D_{\mu},D_{\nu}]$.
From evaluating diagram Fig. \ref{fig:feyn1}b, the factorization formula for three-particle twist-4 contribution reads
\begin{eqnarray}
 F^{3PT4}(Q^2)
 &=&-{\sqrt{2}f_\pi \over 6Q}\int_0^1 {du\over\sqrt{u}}\int_0^\infty
 b^2dbK_1(\sqrt{u}Qb)\rho^{3PT4}(\alpha_i,b),
 \end{eqnarray}
where
\begin{eqnarray}
 \rho^{3PT4}(u,b)
 &=&\int^{\bar u}_0d\alpha_q \int^u_0 {d\alpha_{\bar q}\over
 \alpha_g}\bigg[ {2u-1-\alpha_q-\alpha_{\bar q}\over \alpha_g}\phi^\pi_\parallel
 (\alpha_q,\alpha_{\bar q},1-\alpha_q-\alpha_{\bar q})\nonumber \\ &+&\tilde\phi^\pi_\parallel
 (\alpha_q,\alpha_{\bar q},1-\alpha_q-\alpha_{\bar q})\bigg].
 \end{eqnarray}
Employing the definition \cite{Khodjamirian:1997tk}
\begin{eqnarray}
 \varphi^{TW4}(u,b)=4[g_1(u,b)+G_2(u,b)]+\rho^{3PT4}(u,b),
 \end{eqnarray}
we can write the overall  contribution from twist-4 pion LCDAs as
\begin{eqnarray}
 F^{3PT4}(Q^2)
 &=&-{\sqrt{2}f_\pi \over 6Q}\int_0^1 {du\over\sqrt{u}}\int_0^\infty
 b^2dbK_1(\sqrt{u}Qb)\varphi^{TW4}(u,b),
 \end{eqnarray}
here the Sudakov evolution factor is neglected, because there is no study on the joint resummation effect for the high-twist pion wave function yet.  In this sense,  this result is not complete in $k_T$ factorization framework, but it does not matter because this contribution is actually free from endpoint singularity.

\subsection{Hadronic Structure of Photon}
To investigate the contribution  from the hadronic structure of photon,  the LCDAs of photon \cite{Ball:2002ps} are needed. The definition of two-particle twist-2 and twist-3 LCDAs are given below
\begin{eqnarray}
\langle0|\bar
 q(z)\sigma_{\alpha\beta}q(0)| \gamma(p,\lambda)\rangle
&=&ig_{em}Q_q\langle\bar
qq\rangle(p_\beta\epsilon_\alpha-p_\alpha\epsilon_\beta)\int_0^1
   dx e^{ixp\cdot z}[\chi(\mu)\phi_\gamma(x,\mu)],
\nonumber \\
\langle 0|\bar
 q(z)\gamma_{\alpha}q(0)| \gamma(p,\lambda)\rangle
&=&g_{em}Q_qf_{3\gamma}\epsilon_\alpha\int_0^1
   dx e^{ixp\cdot z}\psi^{(v)}_\gamma(x,\mu),
\nonumber \\
\langle 0|\bar
 q(z)\gamma_{\alpha}\gamma_5q(0)|\gamma(p,\lambda)\rangle
&=&{1\over
4}g_{em}Q_qf_{3\gamma}\epsilon_{\alpha\beta\rho\sigma}p^\rho
z^\sigma\epsilon^{\beta}\int_0^1
   dx e^{ixp\cdot z}\psi^{(a)}_\gamma(x,\mu),
\end{eqnarray}
where $\phi_\gamma(x,\mu)$ is twist-2, and $\psi^{(a,v)}_\gamma(x,\mu)$ are twist-3. At the tree level the trace formulism is convenient to evaluate transition matrix element, so that the following momentum space projector of photon LCDAs is useful
\begin{eqnarray}
M^\gamma_{\alpha\beta} &=&{1\over 4}g_{em}Q_q\bigg\{-\langle\bar
qq\rangle\not\!\epsilon\not\! p\chi(\mu)\phi_\gamma(x,\mu)
+f_{3\gamma}\not\!\epsilon\psi^{(v)}_\gamma(x,\mu)
\nonumber \\
 &-&{i\over 8}f_{3\gamma}\epsilon_{\mu\nu\rho\sigma}(\gamma^{\mu}\gamma^5){ n}^\rho
\epsilon^{\nu}[\bar n^\sigma{d\over
dx}\psi^{(a)}_\gamma(x,\mu)-2E_\gamma\psi^{(a)}_\gamma(x,\mu){\partial\over
\partial k_{\perp\sigma}}]\bigg\}_{\alpha\beta}.
\end{eqnarray}
Similarly, we introduce the momentum space projector of pion wave function up to two-particle twist-3
\begin{equation}
M_{\delta\alpha}^\pi = \frac{i \, f_\pi}{4}
  \, \Bigg\{ \not\! p \gamma_5 \, \phi_\pi(u)  -
   \mu_\pi \gamma_5 \left(  \phi^\pi_p(u)
       - {i\over 2} \sigma_{\mu\nu}\, \bar n^\mu n^\nu \,
       \frac{\phi^{\pi\prime}_\sigma(u)}{6}
 +i \sigma_{\mu\nu}  p^\mu \, \frac{\phi^\pi_\sigma(u)}{6}
       \, \frac{\partial}{\partial
         k_\perp{}_{\nu}} \right)
 \Bigg\}_{\delta\alpha}.
\label{pimeson2}
\end{equation}
Since in PQCD approach the endpoint singularity is regularized, the matrix element of pion transition form factor can be calculated
through the convolution formula
\begin{eqnarray}
\langle \pi|\bar q\Gamma b|\gamma\rangle_{\rm HS}={4\pi\alpha_s
C_F\over N_c}\int_0^1 dx\int_0^\infty b_1db_1\int_0^1
du\int_0^\infty
b_2db_2M^\gamma_{\beta\rho}H^\Gamma_{\alpha\beta\rho\sigma}M^\pi_{\sigma\alpha}.
\end{eqnarray}

\begin{figure}
\begin{center}
\includegraphics[width=0.40 \columnwidth]{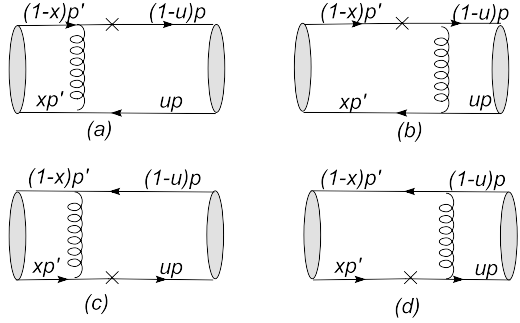} \\
\vspace*{0.1cm} \caption{Fenyman diagrams of contribution of the
pion transition form factor from hadronic structure of photon}
\label{fig:feyn2}
\end{center}
\end{figure}

The hard kernel can be obtained through calculating the Feynman diagrams in Fig \ref{fig:feyn2}, and  the factorization formula
for the form factor reads
\begin{eqnarray}
F^{PHS}(E_\gamma) &=&{4\pi\alpha_sC_Ff_\pi(Q^2_u-Q^2_d)\over
 \sqrt 2N_c}\int_0^1 dx\int_0^\infty b_1db_1\int_0^1
du\int_0^\infty b_2db_2
\nonumber \\
&\times&\bigg
\{h^a_e(x,u,b_1,b_2)f_{3\gamma}\phi_\pi(u)\psi_\gamma^{(v)}(x)+h^b_e(x,u,b_1,b_2)[-f_{3\gamma}x\phi_\pi(u)\psi_\gamma^{(v)}(x)
\nonumber
\\&+&2\chi(\mu)\langle\bar
qq\rangle(\mu)\mu_\pi\phi_\gamma(x)\phi^\pi_p(u)]\bigg\},
\end{eqnarray}
with the PQCD hard function
\begin{eqnarray}
h^a_e(x,u,b_1,b_2)&=&e^{-s_\pi(t)-s_\gamma(t)}\left[\theta(b_1-b_2)I_0(\sqrt
{uQ^2}b_2)K_0(\sqrt {uQ^2}b_1)\right.\nonumber\\
&& \left.+\theta(b_2-b_1)I_0(\sqrt {uQ^2}b_1)K_0(\sqrt {uQ^2}b_2)\right]K_0(\sqrt {xuQ^2}b_1)S_t(u),\nonumber\\
h^b_e(x,u,b_1,b_2)&=&e^{-s_\pi(t)-s_\gamma(t)}\left[\theta(b_1-b_2)I_0(\sqrt
{xQ^2}b_2)K_0(\sqrt {xQ^2}b_1)\right.\nonumber \\
&& \left.+\theta(b_2-b_1)I_0(\sqrt {xQ^2}b_1)K_0(\sqrt
{xQ^2}b_2)\right]K_0(\sqrt {xuQ^2}b_1)S_t(x),
\end{eqnarray}
where the Sudakov factor $s_\gamma(t)$ is the same as that of a vector meson. In this part, because we do not include the NLO QCD corrections, the hard kernel does not depend on the factorization scale, though the form factor is dependent on the factorization scale in principle. In the PQCD approach, in order to suppress the contribution of high order, the factorization scale is set to be  $t=\max(\sqrt xQ, 1/b)$, and we allow an error area of $t$ in the numerical evaluation. Note that the contribution of higher twist LCDAs of photon is also not considered in the present paper, as they are proved to be small in the previous studies \cite{Wang:2018wfj}.
\section{Numerical Analysis}\label{numerical}
The leading twist pion LCDA satisfies the well-known Efremov-Radyushkin-Brodsky-Lepage equation \cite{Efremov:1979qk,Lepage:1980fj}, which indicates that it can be expanded in terms of the Gegenbauer polynomials $ C_n^{3/2}$,
\begin{eqnarray}
\phi_\pi(u, \mu) =  6 u (1-u) \sum_{n=0}^{\infty} a_n(\mu) \,
C_n^{3/2} (2 u -1) \,,
\end{eqnarray}
where the odd Gegenbauer moments $a_{2n+1}$ vanish due to symmetry prosperities. The fundamental ingredients in the LCDA are the even Gegenbauer moments $a_{2n}$. Tremendous efforts have been devoted to the determinations of the lowest moment $a_2(\mu)$ from the calculations with the QCD sum rules \cite{Chernyak:1981zz} and with the lattice simulations, and  by matching  the experimental data. The current widely used models for pion LCDA  include the Bakulev-Mikhailov-Stefanis model \cite{Mikhailov:2016klg,Bakulev:2001pa}, the KMOW model \cite{Khodjamirian:2011ub}, the holographic model \cite{Brodsky:2007hb}, etc. In the present paper we follow \cite{Li:2014xda} to adopt a simpler model with only leading Gegenbauer moment,
\begin{eqnarray}
\phi(u)=  6 \, u (1-u) \, \left[ 1+ a_2 \, C_2^{3/2}(2u-1)\right] \,, \label{initial condition: longitudinal distribution }
\end{eqnarray}
we choose two  values of $a_2$ , i.e., $a_2=0.09$ and $a_2=0.17$, in the numerical evaluation for comparison. The joint resummation improved expression for this model has been obtained, and is given in the appendix. For twist-3 pion LCDAs, we adopt \cite{Li:2012nk}
\begin{eqnarray}
 \phi^\pi_p(u) &=& 1+0.59C^{1\over  2}_2(2u-1)+0.09C^{1\over  2}_4(2u-1)\,\nonumber \\
 \phi^\pi_\sigma(u) &=&6u(1-u)[1+0.11 C_2^{3/2}(2u-1)].
\end{eqnarray}
The twist-4 pion LCDA has the following form if only leading conformal spin contribution is kept \cite{Khodjamirian:1997tk},
\begin{eqnarray}
 \varphi^{TW4}(u,\mu) &=&  {80\over 3}\delta^2_\pi(\mu) \, u^2 (1-u)^2 \,  \,,
\end{eqnarray}
where the normalization parameter is defined by
\begin{eqnarray}
 \langle 0|g_s\bar q\tilde G^{\mu\nu}\gamma_\nu q(0)|\pi(p)\rangle=if_\pi\delta^2_\pi(\mu)p^\mu,
\end{eqnarray}
with the renormalization-scale evolution at one loop
\begin{eqnarray}
\delta_{\pi}^2 (\mu) = \left [ {\alpha_s(\mu) \over
\alpha_s(\mu_0)} \right ]^{32 \over 9  \beta_0}\, \delta_{\pi}^2 (\mu_0) .
\end{eqnarray}
\begin{table}
\centerline{\parbox{14cm}{\caption{\label{para} Numerical value of the parameters in the LCDAs of photon}}} \vspace{0.1cm}
\begin{center}
\begin{tabular}{c|c|c|c}
\hline\hline parameter &$\chi(1GeV)$ & $\langle\bar
qq\rangle(1GeV)$ &$b_2(1GeV)$\\
\hline value &$(3.15\pm 0.03)GeV^{-2}$
&$-[(256^{+14}_{-16})MeV]^3$&$0.07\pm0.07$
\\
\hline\hline parameter  &
$f_{3\gamma}(1GeV)$&$\omega^V_\gamma(1GeV)$ & $\omega^A_\gamma(1GeV)$ \\
\hline value  & $ -(4 \pm 2)
 \times 10^{-3}\,{\rm GeV}^2$ &$3.8\pm 1.8$&$-2.1 \pm 1.0$
\\
\hline\hline
\end{tabular}
\end{center}
\vspace{-0.2cm}
\end{table}
The numerical value of $\delta_{\pi}$ will be taken as $\delta_{\pi}^2(1 \, {\rm GeV})=(0.2 \pm 0.04) \, {\rm GeV^2}$ computed from the QCD sum rules \cite{Novikov:1983jt} (see also \cite{Ball:2006wn}). The light-cone distribution amplitudes $\phi_{\gamma}(u), \psi^{(v,a)}(\omega,\xi)$ have been systematically studied in Ref.\cite{Ball:2002ps}, and the expressions are quoted as follows. The two-particle twist-2 LCDA is expanded in terms of Gegenbauer polynomials,
\begin{eqnarray}
    \phi_\gamma(x,\mu)  &=& 6x\bar x
  \left[1+\sum_{n=2}^\infty b_n(\mu_0)
  C^{3/2}_n(2x-1)\right],
\end{eqnarray}
and twist-3 LCDAs in conformal expansion read
\begin{eqnarray}
\psi^{(v)}(\xi, \mu) &=& 5 \, \left ( 3\, \xi^2 -1  \right ) + {3
\over 64} \, \left [ 15 \, \omega_{\gamma}^{V}(\mu) - 5\,
\omega_{\gamma}^{A}(\mu)  \right ] \,
\left ( 3 - 30 \, \xi^2 +  35 \,  \xi^4 \right ) \,,  \nonumber \\
\psi^{(a)}(\xi, \mu) &=&  {5 \over 2} \, \left ( 1 -  \xi^2 \right
) \, (5 \, \xi^2 -1 ) \left ( 1 + {9 \over 16} \,
\omega_{\gamma}^{V}(\mu) - {3 \over 16} \,
\omega_{\gamma}^{A}(\mu)  \right ) \,.
 \end{eqnarray}
The  value of the parameters used in the LCDAs of photon are presented in Table \ref{para}, among them the scale dependent parameters are given at $\mu_0=1.0GeV$. These parameters should be run to the factorization scale $t$, and the evolution kernel for $\chi(\mu), \langle\bar q q\rangle(\mu), b_2(\mu),f_{3\gamma}(\mu)$ and $ \omega_{V,A}(\mu) $ have been given in \cite{Ball:2002ps}.

\begin{figure}
\begin{center}
\includegraphics[width=0.46 \columnwidth]{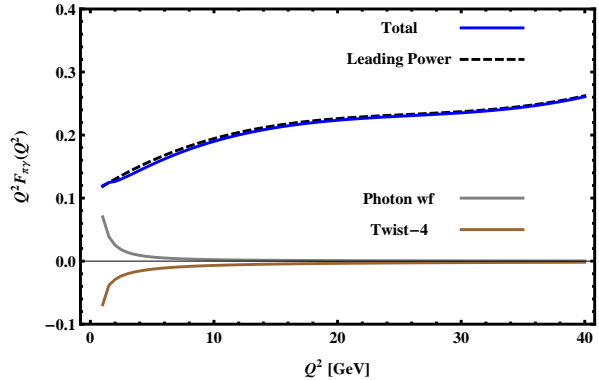} \\
\vspace*{0.1cm} \caption{Contributions of leading power and
subleading power corrections } \label{fig:form1}
\end{center}
\end{figure}
\begin{figure}
\begin{center}
\includegraphics[width=0.46 \columnwidth]{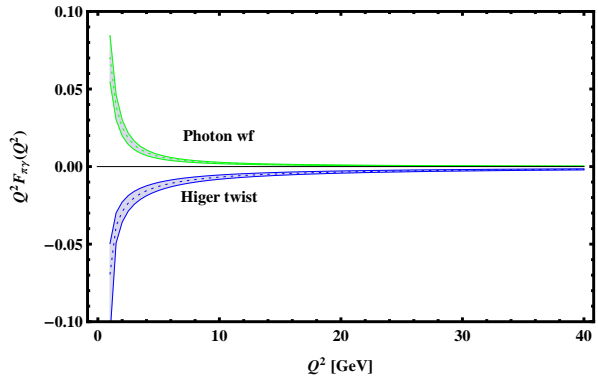} \\
\vspace*{0.1cm} \caption{Uncertainties of the contribution from
subleading power corrections } \label{fig:un}
\end{center}
\end{figure}
Now we turn to investigate the leading power result from joint resummation improved PQCD approach  and the contributions from various sources of subleading power corrections. In Fig. \ref{fig:form1} the result of each kind of contributions are displayed. At the leading power the result from collinear factorization approaches $\sqrt 2f_\pi$ when $Q^2$ goes to infinity, while in $k_T$ factorization with joint resummation, the behavior at large $Q^2$ region is modified. $Q^2F_{\gamma^{\ast} \gamma \to \pi^0} (Q^2)$ slightly increases when $Q^2$ is getting larger, which shows the same tendency with the experimental data. The subleading power corrections start from $1/Q^4$, which fall down rapidly when $Q^2$ increases, and are significant only at small $Q^2$ region. From this figure we can see that the contributions from higher twist pion wave functions and hadronic structure of photon have different sign, the cancellation between them makes the power correction investigated in this paper is minor even at small $Q^2$ region. This result is consistent with the investigation from light-cone sum rules \cite{Wang:2017ijn}. So that the NLP corrections in pion transition form factors is small, which is on the contrary to the  leptonic radiative decay $B\to \gamma \ell \nu$ \cite{Shen:2018abs}, where the NLP corrections decrease the LP result over $50\%$. Our result indicates that the power corrections can hardly modify the leading power prediction from the joint resummation improved PQCD approach, thus the total result  still agrees well with the BaBar data. As a result, the pion transition form factor is a good platform to extract the nonperturbative information on the shape of the leading twist pion distribution amplitude. The uncertainties of the contribution from the higher twist pion wave functions and the hadronic structure of photon are shown in Fig \ref{fig:un}. The main source of the uncertainty is the parameters in the twist-4 Pion LCDA and the LCDAs of photon which are presented in Table.\ref{para}, as well as the factorization scale $t$ which is allowed a float up or down $20\%$ .
\begin{figure}
\begin{center}
\includegraphics[width=0.46 \columnwidth]{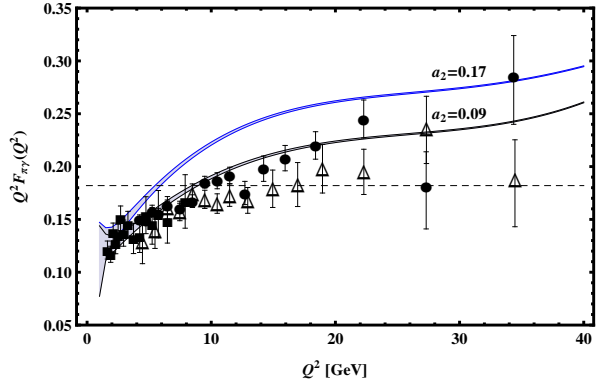} \\
\vspace*{0.1cm} \caption{Comparison between theoretical prediction
and experimental data. The experimental data are from CLEO (squares), BaBar (dots), and Belle (triangles).} \label{fig:exp}
\end{center}
\end{figure}
The comparison between our prediction and the experimental data are presented in Fig. \ref{fig:exp},
where the CLEO \cite{Gronberg:1997fj}, BaBar \cite{Aubert:2009mc} and Belle \cite{Uehara:2012ag} data are shown. The blue and black zones represent the Gegenbauer
moment $a_2=0.17$ and $a_2=0.09$, respectively. At large $Q^2$ region, the pion wave function with $a_2=0.09$
favors the experimental data by Babar and Belle better and the influence from the power corrections are negligible.
 At small $Q^2$ region, the subleading power contribution only changes leading power result slightly,
 and the prediction with $a_2=0.09$ is consistent with the CLEO data. We emphasize that there are large
 theoretical uncertainty when $Q^2<2GeV^2$,  and the perturbation calculation is not reliable at this region,
  thus one needs not take it seriously. We acknowledge that this work is not a systematical study on the
  power corrections within the framework of effective theory, although in the present work the NLP contribution
  is negligible, it is too early to draw the conclusion that the power suppression contribution is not important
   in this process.
\section{Conclusion}
\label{sect:conclusion}
In this paper, within the framework of perturbative QCD approach
based on $k_T$ factorization, we studied the next-leading order of
$\alpha_s$ corrections, the high power corrections, as well as the
joint resummation effect to the pion transition form factor. For
the higher power contributions, we here considered the effects
from higher twist pion wave functions, including two-particle and
three-particle twist-4 wave functions of pion, and the hadronic
structure of photon. In PQCD approach, the transverse momentum of
partons inside pion and ``hadronic" photon regularized endpoint
singularity existing in collinear factorization. The numerical
result indicates that there exists strong cancellation effect
between the two kinds of NLP contribution, thus the NLP
corrections does not change leading power result manifestly. Our
calculations also indicate that the power corrections cannot
explain the anomalous BaBar data. Since there is no confirmed
conclusion on the two experimental results, we hope the Belle-II
experiments can test our result and give the final conclusion in
future. Furthermore, we also note that the pion transition form
factor sharply depend on the leading twist pion LCDA, which can be
used to extract the information of the pion wave function. For
example, a simple pion wave function model with $a_2=0.9$ can be
consistent with the BaBar  data within the uncertainty area. The
power suppressed contributions considered in this paper are only
from some specific sources, to perform a systematical study on
this process we need to analysis the complete NLP operator
\cite{Rothstein:2003wh} base in soft-collinear effective
theory\cite{Bauer:2000yr,Beneke:2002ph}. This work is left for a
future study.

\section*{Acknowledgement}
 This work was supported in part by the National Natural Science Foundation of China under the Grants
 Nos. 11705159, 11447032 and 11575151; and the Natural Science Foundation of Shandong province under
  the Grant No. ZR2018JL001 and ZR2016JL001.



\begin{thebibliography}{99}



\bibitem{Lepage:1980fj}
  G.~P.~Lepage and S.~J.~Brodsky,
  Phys.\ Rev.\ D {\bf 22} (1980) 2157.

\bibitem{Efremov:1979qk}
  A.~V.~Efremov and A.~V.~Radyushkin,
  Phys.\ Lett.\  {\bf 94B} (1980) 245.

\bibitem{Duncan:1979ny}
  A.~Duncan and A.~H.~M\"{u}eller,
  Phys.\ Lett.\  {\bf 90B} (1980) 159.


\bibitem{delAguila:1981nk}
  F.~del Aguila and M.~K.~Chase,
  Nucl.\ Phys.\ B {\bf 193} (1981) 517.




\bibitem{Braaten:1982yp}
  E.~Braaten,
  Phys.\ Rev.\ D {\bf 28} (1983) 524.




\bibitem{Kadantseva:1985kb}
  E.~P.~Kadantseva, S.~V.~Mikhailov and A.~V.~Radyushkin,
  Yad.\ Fiz.\  {\bf 44} (1986) 507
   [Sov.\ J.\ Nucl.\ Phys.\  {\bf 44} (1986) 326].




\bibitem{Melic:2002ij}
  B.~Melic, D.~M\"{u}eller and K.~Passek-Kumericki,
  Phys.\ Rev.\ D {\bf 68} (2003) 014013
  [hep-ph/0212346].



\bibitem{Aubert:2009mc}
  B.~Aubert {\it et al.} [BaBar Collaboration],
  Phys.\ Rev.\ D {\bf 80} (2009) 052002
  [arXiv:0905.4778 [hep-ex]].
\bibitem{Uehara:2012ag}
  S.~Uehara {\it et al.} [Belle Collaboration],
  Phys.\ Rev.\ D {\bf 86} (2012) 092007
  [arXiv:1205.3249 [hep-ex]].



\bibitem{Masjuan:2012wy}
  P.~Masjuan,
  Phys.\ Rev.\ D {\bf 86} (2012) 094021
  [arXiv:1206.2549 [hep-ph]].






\bibitem{Hoferichter:2014vra}
  M.~Hoferichter, B.~Kubis, S.~Leupold, F.~Niecknig and S.~P.~Schneider,
  Eur.\ Phys.\ J.\ C {\bf 74} (2014) 3180
  [arXiv:1410.4691 [hep-ph]].




\bibitem{Gerardin:2016cqj}
  A.~G\'{e}rardin, H.~B.~Meyer and A.~Nyffeler,
  Phys.\ Rev.\ D {\bf 94} (2016)   074507
  [arXiv:1607.08174 [hep-lat]].



\bibitem{Radyushkin:2009zg}
  A.~V.~Radyushkin,
  Phys.\ Rev.\ D {\bf 80} (2009) 094009
  [arXiv:0906.0323 [hep-ph]].




\bibitem{Polyakov:2009je}
  M.~V.~Polyakov,
  JETP Lett.\  {\bf 90} (2009) 228
  [arXiv:0906.0538 [hep-ph]].




\bibitem{Agaev:2010aq}
  S.~S.~Agaev, V.~M.~Braun, N.~Offen and F.~A.~Porkert,
  Phys.\ Rev.\ D {\bf 83} (2011) 054020
  [arXiv:1012.4671 [hep-ph]].

\bibitem{Li:1992nu}
  H.~n.~Li and G.~F.~Sterman,
  Nucl.\ Phys.\ B {\bf 381} (1992) 129.



\bibitem{LM09} H.-n. Li and S. Mishima,
Phys. Rev. D {\bf 80}, 074024 (2009).

\bibitem{Nandi:2007qx}
  S.~Nandi and H.~n.~Li,
  Phys.\ Rev.\ D {\bf 76} (2007) 034008
  [arXiv:0704.3790 [hep-ph]].





\bibitem{Musatov:1997pu}
  I.~V.~Musatov and A.~V.~Radyushkin,
  Phys.\ Rev.\ D {\bf 56} (1997) 2713
  [hep-ph/9702443].




\bibitem{Wu:2010zc}
  X.~G.~Wu and T.~Huang,
  Phys.\ Rev.\ D {\bf 82} (2010) 034024
  [arXiv:1005.3359 [hep-ph]].




\bibitem{Co03}
  J.~C.~Collins,
  Acta Phys.\ Polon.\ B {\bf 34}, 3103 (2003)
  [hep-ph/0304122].
\bibitem{CS08} I. O. Cherednikov and N.G. Stefanis,
Nucl. Phys. {\bf B802}, 146 (2008).

\bibitem{BBDM}
A.~Bacchetta, D.~Boer, M.~Diehl and P.~J.~Mulders,
  JHEP {\bf 0808}, 023 (2008)
  [arXiv:0803.0227 [hep-ph]].

  \bibitem{Collins:2011zzd}
  J.~Collins,
  {\it Foundations of Perturbative QCD },
  Cambridge monographs on particle physics, nuclear physics and cosmology, 32.

\bibitem{Li:2014xda}
  H.~n.~Li and Y.~M.~Wang,
  JHEP {\bf 1506}, 013 (2015)
  doi:10.1007/JHEP06(2015)013
  [arXiv:1410.7274 [hep-ph]].






\bibitem{Li:2013xna}
  H.~N.~Li, Y.~L.~Shen and Y.~M.~Wang,
  JHEP {\bf 1401} (2014) 004
  [arXiv:1310.3672 [hep-ph]].

\bibitem{Khodjamirian:1997tk}
  A.~Khodjamirian,
  Eur.\ Phys.\ J.\ C {\bf 6} (1999) 477
  [hep-ph/9712451].


\bibitem{Wang:2017ijn}
  Y.~M.~Wang and Y.~L.~Shen,
  JHEP {\bf 1712}, 037 (2017)
  doi:10.1007/JHEP12(2017)037
  [arXiv:1706.05680 [hep-ph]].
\bibitem{Agaev:2012tm}
  S.~S.~Agaev, V.~M.~Braun, N.~Offen and F.~A.~Porkert,
  Phys.\ Rev.\ D {\bf 86} (2012) 077504
  [arXiv:1206.3968 [hep-ph]].

\bibitem{Kroll:2010bf}
  P.~Kroll,
  Eur.\ Phys.\ J.\ C {\bf 71} (2011) 1623
  [arXiv:1012.3542 [hep-ph]].



















\bibitem{Nagashima:2002ia}
  M. Nagashima and H.-n. Li,
  Phys.\ Rev.\ D {\bf 67}, 034001 (2003).


\bibitem{Ball:2002ps}
  P.~Ball, V.~M.~Braun and N.~Kivel,
  Nucl.\ Phys.\ B {\bf 649} (2003) 263
  [hep-ph/0207307].

\bibitem{Wang:2018wfj}
  Y.~M.~Wang and Y.~L.~Shen,
  JHEP {\bf 1805}, 184 (2018)
  doi:10.1007/JHEP05(2018)184
  [arXiv:1803.06667 [hep-ph]].



\bibitem{Chernyak:1981zz}
  V.~L.~Chernyak and A.~R.~Zhitnitsky,
  Nucl.\ Phys.\ B {\bf 201} (1982) 492
   Erratum: [Nucl.\ Phys.\ B {\bf 214} (1983) 547].


\bibitem{Mikhailov:2016klg}
  S.~V.~Mikhailov, A.~V.~Pimikov and N.~G.~Stefanis,
  Phys.\ Rev.\ D {\bf 93} (2016)   114018
  [arXiv:1604.06391 [hep-ph]].


\bibitem{Bakulev:2001pa}
  A.~P.~Bakulev, S.~V.~Mikhailov and N.~G.~Stefanis,
  Phys.\ Lett.\ B {\bf 508} (2001) 279
   Erratum: [Phys.\ Lett.\ B {\bf 590} (2004) 309]
  [hep-ph/0103119].



\bibitem{Khodjamirian:2011ub}
  A.~Khodjamirian, T.~Mannel, N.~Offen and Y.-M.~Wang,
  Phys.\ Rev.\ D {\bf 83} (2011) 094031
  [arXiv:1103.2655 [hep-ph]].



\bibitem{Brodsky:2007hb}
  S.~J.~Brodsky and G.~F.~de Teramond,
  Phys.\ Rev.\ D {\bf 77} (2008) 056007
  [arXiv:0707.3859 [hep-ph]].
\bibitem{Li:2012nk}
  H.~n.~Li, Y.~L.~Shen and Y.~M.~Wang,
  Phys.\ Rev.\ D {\bf 85} (2012) 074004
  [arXiv:1201.5066 [hep-ph]].
\bibitem{Novikov:1983jt}
  V.~A.~Novikov, M.~A.~Shifman, A.~I.~Vainshtein, M.~B.~Voloshin and V.~I.~Zakharov,
  Nucl.\ Phys.\ B {\bf 237} (1984) 525.




\bibitem{Ball:2006wn}
  P.~Ball, V.~M.~Braun and A.~Lenz,
  JHEP {\bf 0605} (2006) 004
  [hep-ph/0603063].

\bibitem{Shen:2018abs}
  Y.~L.~Shen, Z.~T.~Zou and Y.~B.~Wei,
  Phys.\ Rev.\ D {\bf 99}, no. 1, 016004 (2019)
  doi:10.1103/PhysRevD.99.016004
  [arXiv:1811.08250 [hep-ph]].


\bibitem{Gronberg:1997fj}
  J.~Gronberg {\it et al.} [CLEO Collaboration],
  Phys.\ Rev.\ D {\bf 57} (1998) 33
  [hep-ex/9707031].


\bibitem{Rothstein:2003wh}
  I.~Z.~Rothstein,
  Phys.\ Rev.\ D {\bf 70} (2004) 054024
  [hep-ph/0301240].


\bibitem{Bauer:2000yr}
  C.~W.~Bauer, S.~Fleming, D.~Pirjol and I.~W.~Stewart,
  Phys.\ Rev.\ D {\bf 63} (2001) 114020
  [hep-ph/0011336].

\bibitem{Beneke:2002ph}
  M.~Beneke, A.~P.~Chapovsky, M.~Diehl and T.~Feldmann,
  Nucl.\ Phys.\ B {\bf 643}, 431 (2002)
  doi:10.1016/S0550-3213(02)00687-9
  [hep-ph/0206152].













\end{thebibliography}
\end{document}